\documentclass[showpacs,amsmath,amssymb,preprint,aps,showkeys]{revtex4}
\usepackage{graphicx}
\usepackage{dcolumn}
\usepackage{bm}
\usepackage{amssymb}
\usepackage{amsmath}
\usepackage{amsthm}
\usepackage{psfrag}

\def \d {\mathrm{d}}
\def \mc {\mathcal}
\def \mb {\mathbb}

\def \G {G\"odel}

\begin{document}
\title{Unwrapping Closed Timelike Curves} \author{Sergei Slobodov}\email{slobodov@phas.ubc.ca}\affiliation{Department of Physics and Astronomy, University of British Columbia, Vancouver, BC, Canada, V6T 1Z1}
\date{\today}
\begin{abstract}
Closed timelike curves (CTCs) appear in many solutions of the Einstein equation, even with reasonable matter sources. These solutions appear to violate causality and so are considered problematic. Since CTCs reflect the global properties of a spacetime, one can attempt to change its topology, without changing its geometry, in such a way that the former CTCs are no longer closed in the new spacetime. This procedure is informally known as unwrapping. However, changes in global identifications tend to lead to local effects, and unwrapping is no exception, as it introduces a special kind of singularity, called quasi-regular. This ``unwrapping'' singularity is similar to the string singularities. We give two examples of unwrapping of essentially 2+1 dimensional spacetimes with CTCs, the Gott spacetime and the {\G} universe. We show that the unwrapped Gott spacetime, while singular, is at least devoid of CTCs. In contrast, the unwrapped {\G} spacetime still contains CTCs through every point. A ``multiple unwrapping'' procedure is devised to remove the remaining circular CTCs. We conclude that, based on the two spacetimes we investigated, CTCs appearing in the solutions of the Einstein equation are not simply a mathematical artifact of coordinate identifications, but are indeed a necessary consequence of General Relativity, provided only that we demand these solutions do not possess naked quasi-regular singularities.
\end{abstract}

\pacs{04.20.Gz}
\keywords{closed timelike curves - G\"odel universe - Gott spacetime - identifications}

\maketitle

\section{\label{sec:intro}Introduction}
Closed timelike curves (CTCs) are closed curves in a spacetime that a timelike test observer can trace \cite{Hawking:1973book}. CTCs are considered problematic as their presence appears to lead to causality violations. Spacetimes with CTCs are usually dealt with in one of two ways: either a spacetime with CTCs is declared not physically relevant or it is modified globally in such a way that the CTCs are absent. In many cases the CTCs are manifest isometries of the spacetime and follow coordinate curves of a periodically identified coordinate. If this periodic identification is removed, the global structure of the spacetime changes. This global modification is informally known as unwrapping. One is then tempted to declare the unwrapped spacetime to be ``more natural'' than the original one. This is the content of the claim of Cooperstock and Tieu in \cite{Cooperstock:2005pu}, who declare that ``the imposition of periodicity in a timelike coordinate is the actual source of CTCs, rather than the physics of general relativity''. We investigate this claim in detail.

Two natural questions that arise are: Do unwrapped spacetimes contains any CTCs not explicitly removed by unwrapping?  Do any other pathologies arise as a result of unwrapping CTCs? To answer these questions we consider the two spacetimes where CTCs are concluded to be artificial by Cooperstock and Tieu, the {\G} universe \cite{Godel:1949ga} and the Gott spacetime \cite{Gott:1990zr}. These spacetimes serve as nice toy models, as they are highly symmetric and essentially $2+1$-dimensional (each is a direct products of a $2+1$-dimensional spacetime with a spacelike real line). The answer to to the first question turns out to be affirmative for the {\G} spacetime, and the answer to the second question is affirmative for both spacetimes studied. 

Implicit in the claim of \cite{Cooperstock:2005pu} seems to be the requirement that the metric of the spacetime without the periodic identifications is locally the same as of the original spacetime with identifications. In particular, if the original spacetime is regular everywhere, so should be the spacetime where the CTCs are unwrapped. We find that in both Gott and {\G} spacetimes there is an obstacle to having this correspondence everywhere, which manifests itself as a singularity in the unwrapped spacetime. This singularity is of the type known in literature as a quasi-regular singularity \cite{Ellis:1977pj}, where the spacetime curvature is bounded along each incomplete curve. 

This ``unwrapping'' singularity is similar to the one describing infinitely thin straight cosmic strings in $3+1$ dimensions, where strings are $1+1$-dimensional singularities. If a $3+1$-dimensional spacetime is a direct product of a $2+1$-dimensional spacetime and a real line, the third spatial dimension can be projected out, transforming the string into a point particle. The main difference between the unwrapping singularity and a point particle singularity is topological: there are no closed curves winding around the unwrapping singularity. 

We show that, while unwrapping the Gott spacetime results in a (singular) spacetime with no CTCs, unwrapping the {\G} space does not remove all of the CTCs. 

In the case of the {\G} spacetime, where the CTCs still persist after unwrapping, we investigate a possibility of ``multiple unwrapping'', where multiple families of CTCs are unwrapped all at once. In this procedure multiple ``strings'' are removed, and the resulting multiply connected spacetime is then unwrapped by constructing its universal cover in order to get rid of the CTCs winding around each removed string. This removes all of the circular CTCs winding around each such string and we conjecture that no other CTCs remain, either.

This paper is organized as follows. In section \ref{sec:spacetimes-with-ctcs} we give a brief review of some of the spacetimes admitting CTCs and describe how the {\G} and Gott solutions fit into the picture. In section \ref{sec:unwrapping} we define what we mean by unwrapping and we investigate the nature of the resulting singularities. In section \ref{sec:gott} we unwrap the Gott spacetime and show that there are no CTCs in the unwrapped (singular) spacetime. In section \ref{sec:godel-unwrap} we unwrap the {\G} space and show that the unwrapped space is singular, and moreover, that CTCs are still present. In section \ref{sec:godel-unwrap-multi} we improve the unwrapping procedure in order to remove the remaining circular CTCs and discuss the properties of the resulting space.  

\section{\label{sec:spacetimes-with-ctcs}Spacetimes with CTCs}
Spacetimes with CTCs can arise in a variety of ways. In some cases, such as the van Stockum cylinder, the {\G} universe and the Kerr blackhole, CTCs are produced by the the ``frame dragging'' effect of the rotating matter. In other cases, such as the spinning string, they are due to coordinate identifications. In yet other cases the CTCs arise due to the non-trivial topology of the spacetime itself (wormholes). We give a brief overview of some of these spacetimes in this section, with the emphasis on the {\G} spacetime.

The first spacetime where the CTCs are manifest, the van Stockum cylinder, was constructed by Lanczos in 1924, then rediscovered by van Stockum \cite{vanStockum:1937} and analyzed by Tipler \cite{Tipler:1974gt}. This spacetime is stationary and axisymmetric (it admits two commuting Killing vectors, one timelike and one spacelike with closed orbits), its metric is of the Weyl-Papapetrou type \cite{Stephani:2003tm}:
\begin{equation}
	\label{eq:vanstockum}
	ds^2=-A(r)dt^2+B(r)dtd\phi+C(r)d\phi^2+H(r)(dr^2+dz^2).
\end{equation}
Here $\phi$ is $2\pi$-periodic, and so the CTCs appear whenever $C(r)<0$. Conventionally, the solution consists of a spinning dust cylinder matched to an external vacuum solution. For certain values of the cylinder size and angular momentum the CTCs occur in the external vacuum only. See \cite{Tipler:1974gt} for details.

One of the most famous solutions of the Einstein equation which admit CTCs was obtained in 1949 by {\G} (\cite{Godel:1949ga}). Its geodesics were computed in \cite{Chandra:1961} and its properties are discussed in \cite{Hawking:1973book}. Following \cite{Hawking:1973book} section 5.7, we write the metric of the {\G} universe as 
\begin{equation}
	\label{eq:godel-cartesian}
	ds^2=-dt^2+dx^2-\frac 12 e^{2\sqrt{2} \omega x}dy^2-2e^{\sqrt{2}\omega x}dtdy + dz^2.
\end{equation}
Here $\omega=\textrm{const}$ and $(t,x,y,z)$ take all real values. In the following we will call this coordinate system Cartesian. The manifold of the {\G} metric is $\mb{R}^4$ and the spacetime is homogeneous. The matter source in this space can be written as 
\begin{equation}
	\label{eq:godel-tab}
	T_{ab}=\rho u_au_b+\frac 12 \rho g_{ab},
\end{equation}
where $\rho=2\omega^2$ is the energy density in the units where $8\pi G=1$ and $c=1$, so that the Einstein equation in the presence of the cosmological constant reads $G_{ab}+\Lambda g_{ab}=T_{ab}$. We use this convention throughout this paper. Here $u_a$ is the timelike unit vector field tangent to the coordinate curves of $t$ in (\ref{eq:godel-cartesian}). If the second term in (\ref{eq:godel-tab}) is associated with a negative cosmological constant $\Lambda=-\frac 12 \rho$, then the matter content of the {\G} universe is rotating dust (pressureless perfect fluid) with density $\rho$, and $\omega$ is the magnitude of its vorticity flow. This spacetime is a direct product of a three-dimensional spacetime with a real line $\mb{R}$, parameterized by the coordinate $z$, which does not add any interesting features and can be safely ignored. In the following we set $\omega=1/\sqrt{2}$, so that $\rho=1$. This is equivalent to rescaling the coordinates, up to a constant overall factor in the metric. 

The {\G} space is highly symmetric, admitting 4 out of a possible 6 Killing vectors in the $(t,x,y)$ subspace. These are $\partial_t,\partial_y,\partial_x-y\partial_y,-2e^{-x}\partial_t+y\partial_x+(e^{-2x}-y^2/2)\partial_y$. The first one of these commutes with the rest, the last three form an $SO(2,1)$ Lie algebra.

The rotating matter in the {\G} space leads to CTCs, some of which are manifest after a coordinate transformation to a cylindrical-like chart $(\tau,r,\phi)$, where $r>0$, $\tau$ can take any real values and $\phi$ is a $2\pi$-periodic angular coordinate:
\begin{equation}
\begin{split}
	\label{eq:godel-cylind}
	e^{x}&=\cosh2r+\cos\phi \sinh 2r \\
	ye^{x}&=\sqrt{2}\sin\phi \sinh 2r \\
	\tan\frac 12 \left(\phi+\frac{t}{\sqrt{2}} -\sqrt{2}\tau\right)&=e^{-2r}\tan\frac 12 \phi.
\end{split}
\end{equation}
The new metric, after omitting the irrelevant z-coordinate, is 
\begin{equation}
	\label{eq:godel-cyl}
	ds^2=-d\tau^2+dr^2+\sinh^2r\left(1-\sinh^2r\right)d\phi^2-2\sqrt{2}\sinh^2{r}d\tau d\phi.
\end{equation}
The validity of imposing periodicity on $\phi$ follows from the fact that the last of the equations (\ref{eq:godel-cylind}) is $2\pi$-periodic in $\phi$, and from the regularity condition on the axis. Specifically, a spacetime admitting an axial ($U(1)$) Killing vector $\xi^a$, parameterized by a $2\pi$-periodic coordinate $\phi$ is regular on the rotation axis (a set of fixed points of $\xi^a$) if and only if the following condition holds:
\begin{equation}
	\label{eq:regularity}
	\frac{(\nabla_a{(\xi^c\xi_c)})(\nabla^a{(\xi^c\xi_c)})}{4\xi^c\xi_c}\to 1,
\end{equation}
where the limit corresponds to the rotation axis (\cite{Stephani:2003tm}). This condition holds on the axis $r=0$ of (\ref{eq:godel-cyl}). The tangent to the coordinate curve of $\phi$ is future pointing when it is timelike, resulting in CTCs. This occurs for $\sinh{r}>1$. 

The {\G} spacetime has generally been discarded as pathological, however other CTC-admitting solutions have been harder to dismiss.

There are several known examples of non-simply connected spacetimes with matter sources satisfying the Null, Weak and Dominant energy conditions \footnote{The Null Energy Condition holds if $T_{ab}k^ak^b\ge 0$ for all null $k^a$, the Weak Energy Condition holds if $T_{ab}k^ak^b\ge 0$ for all timelike $k^a$, and the Dominant Energy Condition holds if $T_{ab}k^a$ is future-pointing for all non-spacelike $k^a$. One or more of these conditions are satisfied by all known classical matter sources.\label{fn-energy-cond}} where CTCs are present. In such cases Carter, in his investigation of the spinning black hole metric (\cite{Carter:1968rr}), makes a distinction between the ``trivial'' CTCs (those that are not homotopic to zero, i.e. non-contractible) and the ``non-trivial'' (contractible) ones, such as those present in the {\G} spacetime. The trivial CTCs can be removed by going from a given non-simply connected spacetime to its universal cover, without changing the metric locally. The non-trivial ones obviously persist even in that case. 

A simple example of a spacetime admitting trivial CTCs is the Minkowski spacetime 
\begin{equation}
	\label{eq:mink-period}
	ds^2=-dt^2+dx^2+dy^2+dz^2
\end{equation}
with the periodically identified timelike coordinate ($t\sim t+T$). This spacetime is homeomorphic to a cylinder $\mb{S}\times\mb{R}^3$, all CTCs are trivial and can be removed by going to its universal covering space, the usual $\mb{R}^4$ manifold with Minkowski metric.

Another example, relevant to some of the unwrapping constructs below, is the spacetime of an infinite spinning cosmic string. A single straight non-spinning cosmic string along the $z$-direction is described by the metric
\begin{equation}
	\label{eq:string}
	ds^2=-dt^2 + dr^2+\left(1-\frac m {2\pi}\right)^2r^2d\phi^2+dz^2,
\end{equation}
$0\le\phi\le{2}\pi$, $m\ne{0}$. This was first analyzed by Marder \cite{Marder:1959}, who described the conical singularity at $r=0$, without assigning any physical meaning to it. For the spinning string the metric is
\begin{equation}
	\label{eq:spin-string}
	ds^2=-\left(dt+\frac a {2\pi} d\phi\right)^2 + dr^2+\left(1-\frac m {2\pi}\right)^2r^2d\phi^2+dz^2,
\end{equation}
$0\le\phi\le 2\pi$, where $a$ is the angular momentum per unit length, and $m$ is the mass per unit length (string tension), respectively. This spacetime can be obtained from that of a non-rotating string (\ref{eq:string}) by replacing the usual identification $(t,r,\phi,z)\sim(t,r,\phi+2\pi,z)$ with $(t,r,\phi,z)\sim(t+a,r,\phi+2\pi,z)$ and substituting $t\to t+a\phi/2\pi$ (see e.g. \cite{Deser:1983tn}). This is an example of ``topological frame dragging'', where an observer on the Killing horizon (corresponding to the zeros of the norm of $\partial_\phi$) appears to be rotating relative to an observer at infinity (\cite{Herrera:1997wi}, \cite{Culetu:2006cc}). The manifold of (\ref{eq:spin-string}) is regular and flat everywhere except at $r=0$, where there is a conical singularity due to the mass term (the deficit angle is equal to $m$). The coordinate curves of $\phi$ are closed and become timelike sufficiently close to the string ($r<\frac a{2\pi-m}$). The conical singularity can be smeared out by a suitable matter distribution \cite{Jensen:1992wj}, in which case the CTCs become contractible.

Instead of a single rotating string one can produce CTCs with a pair of non-rotating strings moving with respect to each other with a non-zero impact parameter, as discovered by Gott \cite{Gott:1990zr}. The coordinate chart in the vicinity of each string is just (\ref{eq:string}), and the two charts can be smoothly connected, such that there is a boost along the junction, as discussed in detail in section \ref{ssec:gott-constr}. This causal structure of this spacetime is, in a sense, an opposite of the spinning string one, as the CTCs in it have minimum size and extend all the way to infinity \cite{Ori:1991ja}. The CTCs of this spacetime are discussed in detail in section \ref{ssec:gott-ctcs}. 

In all of the above examples the CTCs exist for all times, and so such spacetimes are usually considered unphysical, since they do not admit a Cauchy surface from which such a spacetime could evolve.

Another well-known example of a vacuum spacetime with CTCs is the Kerr black hole. In this metric there exist both the CTCs that wrap around the ring singularity and those that do not (\cite{Carter:1968rr}). Both kinds are hidden from an external observer by the event horizon of the black hole.

A different class of spacetimes admitting CTCs are those with non-trivial topology (wormholes) and ``exotic'' matter sources (those violating the energy conditions), but without singularities. Exotic matter is required to keep the wormholes traversable. The first traversable wormhole metric was given in \cite{Morris:1988tu}. An example of such a metric is
\begin{equation}
	\label{eq:wormhole}
	ds^2=-dt^2 + dr^2+\left(r^2+R\left(r\right)^2\right)\left(d\theta^2+\sin^2\theta\d\phi^2\right),
\end{equation}
where $R(r)$ has compact support, $r\in\mb{R}$. Positive and negative values of $r$ correspond to two different flat asymptotic regions. These can be patched together far enough from the wormhole throats, where $R(r)=0$. To do that we can identify a 3-plane (e.g. $x=r\sin\theta\cos\phi=\textrm{const}$) in one of these asymptotic regions with a corresponding 3-plane in the other. Since the spacetime is flat there, both the intrinsic and extrinsic curvatures of the planes vanish, and so a spacetime with such an identification satisfies the Einstein equation without the need for any additional matter sources. Once a wormhole exists in a given spatial slice $t=\textrm{const}$, it can be manipulated into producing CTCs by a variety of means, all based on changing the relative rates of time flow between the throats, as seen by an asymptotic observer, until there is a CTC threading through them. This can be achieved using either the special relativistic time dilation effect, as described in the original paper, or the gravitational one \cite{Frolov:1990si}. Since all such CTCs result from the underlying spacetime not being simply connected, going to the simply connected universal cover gets rid of the CTCs.

Another evidence of the ubiquity of CTCs is the solution constructed by Ori \cite{Ori:2007kk}, where a regular Cauchy horizon bounded by a closed null geodesic develops from a regular spatial slice and the matter sources satisfying energy conditions.

There is a number of conjectures and theorems that deal with the CTCs and their appearance. Tipler \cite{Tipler:1976bi} has shown that CTCs cannot evolve from non-singular initial data in a regular asymptotically flat spacetime. Hawking \cite{Hawking:1991nk} has advanced the Chronology Protection Conjecture, which states that, if the Null Energy Condition holds, then the Cauchy horizon (the null boundary of the domain of validity of the Cauchy problem, see e.g.\cite{Hawking:1973book}) cannot be compactly generated. Moreover, even if the Null Energy Condition is violated, the quantum effects are likely to prevent the Cauchy horizon from appearing. 

\section{\label{sec:unwrapping}Unwrapping}
Since one of our goals is to show that unwrapping contractible CTCs creates singularities, we first review the definition of the singularity in a manifold without boundary, as defined by Schmidt in \cite{Schmidt:1971uf}. Next we discuss the quasi-regular singularities resulting from changing the angular coordinate identifications in Minkowski spacetime, as described in \cite{Ellis:1977pj}.

\subsection{\label{ssec:singularity}Singular Spacetimes}
In General Relativity a spacetime $(\mc{M},g)$ is called singular when it is incomplete, i.e. when it has incomplete inextendible curves \cite{Hawking:1973book}. An incomplete curve is defined as a map $C:[0,1)\to\mc{M}$. Schmidt \cite{Schmidt:1971uf} has developed a way to determine whether the curve ends at a finite distance, by generalizing the notion of affine distance along geodesics to arbitrary curves. The limit points of the incomplete curves define what is known as the ``b-boundary''. See e.g. \cite{Ellis:1977pj} for details. Once the b-boundary is determined, one can talk about a ``neighborhood of a singularity'' and the behavior of tensors, such as the Riemann tensor, along a curve approaching the singularity. 

If a b-boundary point can be included in the interior of some extension of $\mc{M}$, then this point is called a regular b-boundary point, otherwise it is called a singular b-boundary point \cite{Ellis:1977pj}. In the case of singularities constructed by changing the periodic coordinate identifications, the b-boundary coincides with the set of the fixed points of the periodic coordinate before the identification is changed. For example, for the two-dimensional cone with the metric $ds^2=dr^2+r^2d\phi^2$, where $0\le\phi\le\Phi<2\pi$, $r=0$ is its singular b-boundary, due to the deficit angle resulting in the violation of the regularity condition (\ref{eq:regularity}).

It is worth noting that the b-boundary is different from the usual boundary of an $n$-dimensional manifold with a boundary, although there is some overlap. The latter is itself an $(n-1)$-dimensional manifold without boundary. In contrast, the b-boundary can be of any dimension, it may or may not be a manifold itself, and may or may not have a boundary (or a b-boundary). See \cite{Ellis:1977pj} for examples and counter-examples. A conical singularity in the 4-dimensional spacetime with the metric (\ref{eq:string}) is an example of a two-dimensional b-boundary which is a flat $\mb{R}^2$ manifold. The $r=0$ singularity of the Schwarzschild spacetime can be described as a singular b-boundary with a rather peculiar structure (\cite{Johnson:1977dd}).

\subsection{\label{ssec:quasi-regular}Quasi-regular Singularities in flat spacetime}
The singularities arising from unwrapping are not associated with divergent curvature. In the classification of \cite{Ellis:1977pj} these are called quasi-regular. In this case all components of the Riemann curvature tensor in an orthonormal frame parallelly propagated along a curve are bounded for all incomplete curves with an endpoint at the singularity. The other types are: the scalar curvature singularities, where at least some of the curvature invariants, such as $R$ or $R_{ab}R^{ab}$ grow unbounded along an incomplete curve; and the parallelly propagated curvature singularities, where at least some of the components of the Riemann tensor cannot be bounded along some incomplete curves, even though all of the curvature invariants remain finite or even vanish, such as in the case of singularities formed by gravitational waves. 

Unwrapping does not change the metric at any of the regular points, so the unwrapping singularities are always quasi-regular. Provided the original spacetime is asymptotically flat and has no event horizon, neither does the unwrapped one, so the resulting quasi-regular singularity is naked, i.e. there are future directed null curves originating arbitrarily close to the singularity that reach the future null infinity. We first describe the unwrapping singularity for the 4-dimensional Minkowski spacetime and then generalize the definition to apply to the spacetimes of interest. 

\begin{figure}
	\centering
	\psfrag{phi0}{\tiny{$0\le\phi\le 2\pi$}}
	\psfrag{phi1}{\tiny{$0\le\phi\le 2\pi$}}
	\psfrag{phi2}{\tiny{$0\le\phi\le 2\pi$}}
	\psfrag{phi3}{\tiny{$0\le\phi\le 2\pi$}}
	\psfrag{bphi0}{\tiny{$-2\pi\le\bar{\phi}\le 0$}}
	\psfrag{bphi1}{\tiny{$0\le\bar{\phi}\le 2\pi$}}
	\psfrag{bphi2}{\tiny{$2\pi\le\bar{\phi}\le 4\pi$}}
	\psfrag{bphi3}{\tiny{$4\pi\le\bar{\phi}\le 6\pi$}}
	\includegraphics[width=1.0\textwidth]{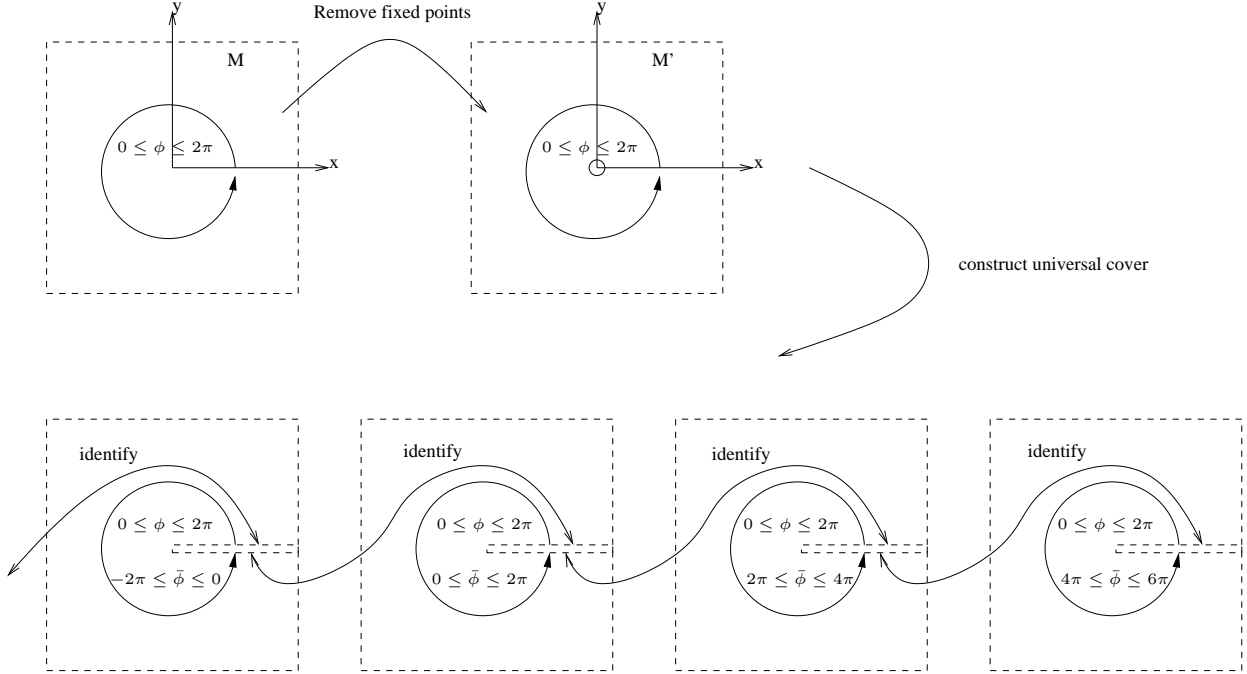}
	\caption{Unwrapping Minkowski space: a) pick a cylindrical chart, b) remove the fixed points of $\phi$ at $r=0$, c) go to the universal cover by extending the range of $\phi$ to all reals. A $t=z=0$ slice is shown as a collection of Cartesian charts with $y=0,x\ge 0$ removed and the surfaces $\phi=2\pi$ of one chart identified with the surface $\phi=0$ of the next chart. $\bar{\phi}$ is the global angular coordinate, $\bar{\phi}\in\mb{R}$.}
	\label{fig:mink-unwrapped}
\end{figure}

\begin{figure}
	\centering
		\includegraphics[width=0.5\textwidth]{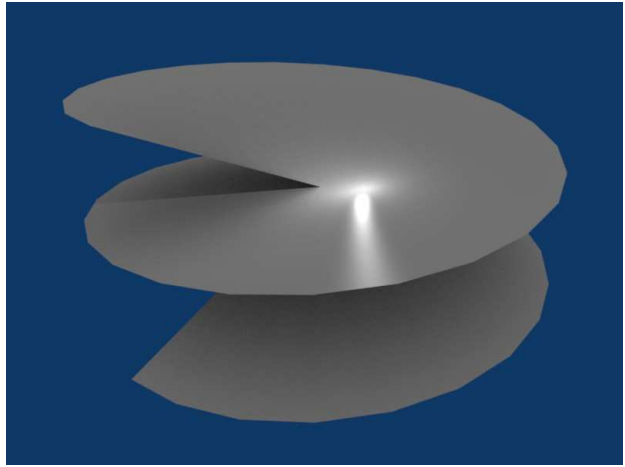}
	\caption{A visualization of the two-dimensional spatial slice of the unwrapped Minkowski space.}
	\label{fig:helix}
\end{figure}

A conical singularity in a 4-dimensional Minkowski spacetime can be obtained as follows (\cite{Ellis:1977pj}):
\begin{enumerate}
\item \label{cone:remove} Remove the timelike two-plane $x=y=0$ in the Cartesian chart. This is precisely where the cylindrical chart is not defined ($r^2=x^2+y^2=0$). This space is no longer simply connected and has the topology of $\mb{S}^1\times\mb{R}^3$. 
\item{\label{cone:unwrap}}Unwrap the resulting space to obtain its universal covering $(\bar{\mc{M}},\bar{g})$, with the same flat metric in the cylindrical chart, but with the range of the new angular coordinate $\bar{\phi}$ extended to $\bar{\phi}\in\mb{R}$, instead of $0\le\phi\le 2\pi$. A local Cartesian chart (with the non-negative $x$-axis removed) is obtained by the standard transformation $(x=r\cos\phi,y=r\sin\phi)$. A two-dimensional slice of this spacetime in the $x-y$ plane ($t=z=0$) is shown in Fig. \ref{fig:mink-unwrapped} and the three-dimensional embedding of this slice into $\mb{R}^3$ is shown in Fig. \ref{fig:helix}. A well-known version of this space is the Riemann surface forming the domain of the complex Log function.
\item{\label{cone:identify}}Identify the points under translation through an angle $\Phi\ne 2\pi$, taking care to preserve the rotational isometry $r=\textrm{const}$, i.e. $(t,r,\phi,z)\sim(t,r,\phi+\Phi,z)$. 
\end{enumerate}

For $\Phi\ne2\pi$ we get the metric of the cosmic string (\ref{eq:string}) with string tension $m$ equal to the deficit angle $m=2\pi-\Phi$. If $\Phi>2\pi$, the string has negative tension. The presence of the conical singularity is reflected in the focusing (or defocusing) of the geodesics passing on the opposite sides of the singularity. This can also be seen from the violation of the regularity condition (\ref{eq:regularity}). This condition is obviously satisfied for the ordinary Minkowski spacetime, where $\xi^c\xi_c=r^2$ in the cylindrical chart, but not after the identifications where the deficit angle $m\ne{0}$, as can be seen by a coordinate transformation where the deficit angle is traded for a constant factor in $\xi^c\xi_c$, such as in (\ref{eq:string}), breaking (\ref{eq:regularity}).

If the last step is omitted, we get the unwrapped space $(\bar{\mc{M}},\bar{g})$, and the singularity is not conical in the usual sense, as there is no closed curve wrapping around it. We will call this type of quasi-regular singularity the unwrapping singularity. 

The subspace $x=y=r=0$ removed in the step \ref{cone:remove} forms the b-boundary of each of the three spaces corresponding to the three steps above. After step \ref{cone:remove} the b-boundary is regular, since it can be included back in to form the original inextendible Minkowski spacetime. 

After step \ref{cone:unwrap} the b-boundary is no longer regular. Indeed, if it were regular, we would be able to include it back into the spacetime and show that any neighborhood of a b-boundary point is homeomorphic to an open ball in $\mb{R}^4$. This would in turn imply that there are closed curves around any such point. Furthermore, any such closed curve is homotopic to a closed curve $r=\textrm{const}$ lying inside the open ball. However, step \ref{cone:unwrap} explicitly removes all such curves, contradicting the regularity assumption. The singularity is quasi-regular, because the Riemann tensor vanishes for all points $r>0$. 

After the step \ref{cone:identify} the b-boundary is still singular, as any arbitrarily small curve around it has the same fixed deficit angle, breaking the regularity condition (\ref{eq:regularity}).

The argument that the b-boundary of the unwrapped spacetime is quasi-regular applies to a class of regular spacetimes that is larger than just the Minkowski space. For the steps \ref{cone:remove} and \ref{cone:unwrap} to be applicable for a certain spacetime, it is sufficient to have the spacetime covered by a single Cartesian chart. In particular, it is valid for the Gott and {\G} spacetimes. Moreover, for any future-directed null curve connecting the origin point $O$ in this Cartesian chart with a certain point $P$ in the original spacetime there is a future-directed null curve connecting the b-boundary in the unwrapped spacetime with one of the infinitely many copies of the point $P$ resulting from unwrapping. Thus, if the unwrapping point is not hidden by an event horizon in the original asymptotically flat spacetime, the quasi-regular singularity in the unwrapped spacetime is necessarily naked.

For completeness, we mention another type of quasi-regular singularity, the Misner singularity, constructed by removing a spacelike two-plane $t=z=0$ from Minkowski space, then periodically identifying points under a given boost $t^2-z^2=\textrm{const}$, to obtain the 4-dimensional Misner space (a direct product of the two-dimensional Misner space introduced in \cite{Misner(1967)} with the $x-y$ plane). The topology of the resulting space is $\mb{S}\times\mb{R}^3$, it contains CTCs through every point, and the surface $t=z=0$ is a quasi-regular singularity. See \cite{Ellis:1977pj} for detailed examples. If we omit the identification step, we obtain an ``unwrapped'' Misner space. 

One can construct rather complicated quasi-regular singularities by cutting and gluing together spacetime pieces with different properties. For example, Krasnikov \cite{Krasnikov:2006zx} describes string-like singularities that are loops or spirals.

\section{\label{sec:gott}Unwrapping Gott spacetime}
The Gott spacetime, discussed in detail in this section, admits CTCs. Cooperstock and Tieu suggested that such a matching is artificial and that identification ``before the Lorentz boost is applied'' is ``more natural'' (\cite{Cooperstock:2005pu}). Since their claim is based on a different spacetime (they appear to remove two timelike ribbons from Minkowski space and boost the resulting singularities relative to each other) and relies on a closed curve crossing these ribbon singularities, it is hard to evaluate. Instead, in keeping with the procedure described in section \ref{ssec:quasi-regular}, we unwrap the Gott spacetime, such that the Gott CTCs correspond to open curves in the unwrapped spacetime. To do that without changing the metric locally, we remove a timelike line from the Gott spacetime and construct the universal cover of the resulting multiply connected (and now singular) spacetime. We show that no CTCs are present in the new spacetime. We also construct some alternative extensions to the Gott spacetime with a timelike line removed and discuss their properties.

\subsection{\label{ssec:gott-constr}Construction}
\begin{figure}
	\centering
	\includegraphics[width=0.9\textwidth]{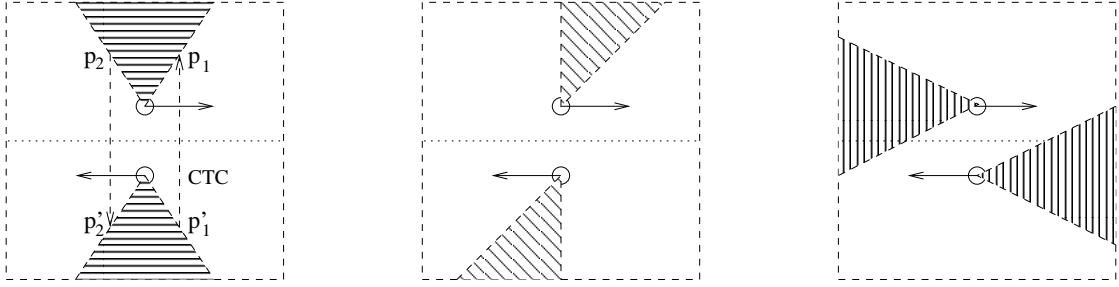}
	\caption{Conical wedge identification choices in the Gott spacetime. Wedge fill lines indicate the identified points. The strings are shown at the moment of the closest approach. Left: Gott identification, where CTCs are manifest (one CTC is shown). Center: Cutler identification, used to prove the existence of points not lying on any CTCs. Right: Carroll identification, used to visualize the existence of CTC-free spacelike hypersurfaces, as the opposite sides of the wedge are identified at equal times.}
	\label{fig:gott-ident}
\end{figure}

We first review the way the $(2+1)$-dimensional Gott spacetime is constructed. Following Gott \cite{Gott:1990zr}, we start with the $(2+1)$-dimensional version of the straight cosmic string spacetime (\ref{eq:string}), where the string is represented by a point particle:
\begin{equation}
	\label{eq:string2d}
	ds^2=-dt^2 + dr^2+\left(1-\frac m {2\pi}\right)^2r^2d\phi'^2,
\end{equation}
where $0\le\phi'\le{2}\pi$ and the particle mass $m$ is the deficit angle. The deficit angle can be made explicit by the substitution $\phi'=\phi/\left(1-\frac m {2\pi}\right)$, where now $0\le\phi\le 2\pi-m$:
\begin{equation}
	\label{eq:string2d-flat}
	ds^2=-dt^2 + dr^2+r^2d\phi^2.
\end{equation}
The coordinate identification $\phi\sim\phi+2\pi-m$ corresponds to removing a timelike wedge centered at the particle and identifying the opposite faces of the wedge. The angle $\phi_0$ the wedge makes with the horizontal axis corresponds to the remaining coordinate gauge freedom and can be chosen in a way that simplifies a particular calculation. Gott has chosen the wedge angle in a way that identifies the surfaces $\phi_0=\frac{\pi}{2}-\frac{m}{2}\sim\phi_1=\frac{\pi}{2}+\frac{m}{2}$, which simplifies his proof of the existence of CTCs. Cutler \cite{Cutler:1992nn} used the identification $\phi_0=\frac{\pi}{2}-m\sim\phi_1=\frac{\pi}{2}$ for one of the strings to show the existence of a timelike cylinder enclosing both strings which no CTCs enter. Carroll et al. \cite{Carroll:1994hz} identified $\phi_0=-\frac{m}{2}\sim\phi_1=\frac{m}{2}$ to visualize the existence of spacelike hypersurfaces through which no CTCs pass. These identification choices are illustrated on Fig. \ref{fig:gott-ident}. As our goal is to investigate the CTCs in the Gott spacetime, we use the Gott's choice of $\phi_0$.

We write each face of the wedge (denoted by the indices 1 and 2) in the Cartesian coordinate system, expressed in terms of two parameters $t$ and $x$:
\begin{equation}
\begin{split}
	\label{eq:gott-wedge-cartesian}
	t_1&=t, \\
	x_1&=x, \\
	y_1&=x\cot{m/2}, \\
	t_2&=t, \\
	x_2&=-x, \\
	y_2&=x\cot{m/2},
\end{split}
\end{equation}
Points corresponding to the same values of $(t,x)$ on both faces are identified. 

It is possible to have multiple strings in the same spacetime, as long as the total deficit angle does not exceed $2\pi$, otherwise the topology of the spacetime becomes $\mb{S}^2\times\mb{R}$ instead of $\mb{R}^3$, where the total deficit angle is equal to $4\pi$ \cite{Deser:1983tn}.

Next step is to boost the wedge in the positive $x$ direction with the velocity $v<c=1$. The coordinates of the faces are then Lorentz-transformed into the laboratory frame as $t_L=\gamma(t+vx)$, $x_L=\gamma(x+vt)$, $y_L=y$:
\begin{equation}
\begin{split}
	\label{eq:gott-wedge-cartesian-boosted}
	t_{1L}&=\gamma(t+vx), \\
	x_{1L}&=\gamma(x+vt), \\
	y_{1L}&=x\cot{m/2}, \\
	t_{2L}&=\gamma(t-vx), \\
	x_{2L}&=\gamma(-x+vt), \\
	y_{2L}&=x\cot{m/2}.
\end{split}
\end{equation}
One can see that the identified points $p_1=(t_{1L},x_{1L},y_{1L})$ and $p_2=(t_{2L},x_{2L},y_{2L})$ have different values of the time coordinate $t_L$ in the laboratory (center of momentum) frame. Specifically, the time difference between the two is
\begin{equation}
	\label{eq:gott-jump}
	\Delta t_L=t_{2L}-t_{1L}=-2\gamma x=v(x_{2L}-x_{1L}),
\end{equation}
and is always negative.
To describe two boosted strings of masses $m$ moving along the $x$-axis in opposite directions with the velocities $v$ and $-v$ and the impact parameter $2b$, we shift the boosted wedge (\ref{eq:gott-wedge-cartesian-boosted}) by $b$ in the positive $y$ direction and introduce a second wedge, $v\to-v'$ and $y\to-y$: 
\begin{equation}
\begin{split}
	\label{eq:gott-full}
	t_{1L}&=\gamma(t+vx), \\
	x_{1L}&=\gamma(x+vt), \\
	y_{1L}&=b+x\cot{m/2}, \\
	t_{2L}&=\gamma(t-vx), \\
	x_{2L}&=\gamma(-x+vt), \\
	y_{2L}&=b+x\cot{m/2}, \\
	t_{1L}'&=\gamma(t'-vx'), \\
	x_{1L}'&=\gamma(x'-vt'), \\
	y_{1L}'&=-b-x'\cot{m/2}, \\
	t_{2L}'&=\gamma(t'+vx'), \\
	x_{2L}'&=\gamma(-x'-vt'), \\
	y_{2L}'&=-b-x'\cot{m/2}, \\
\end{split}
\end{equation}
where primed variables describe the second wedge. We have now constructed the Gott spacetime in a single Cartesian chart $(t_L,x_L,y_L)$, corresponding to the laboratory frame, subject to the two wedge identifications $(t_{1L},x_{1L},y_{1L})\sim(t_{2L},x_{2L},y_{2L})$ and $(t_{1L}',x_{1L}',y_{1L}')\sim(t_{2L}',x_{2L}',y_{2L}')$. If we choose $t'$ and $x'$ such that $\gamma(t+vx)=\gamma(t'-vx')$ and $\gamma(x+vt)=\gamma(x'-vt')$, then the closest approach of the strings corresponds to $t_L=0$.
\subsection{\label{ssec:gott-ctcs}CTCs of the Gott spacetime}
Following Gott, we now consider the curves composed of two pieces of geodesics with $x_L=\textrm{const}$, as shown on the Fig. \ref{fig:gott-ident} left. The first piece connects the two wedges at the points $p_1'=(t_{1L}'=-T,x_{1L}'=a,y_{1L}'=-VT)$ and $p_1=(t_{1L}=T,x_{1L}=a,y_{1L}=VT)$. Here $V$ is the velocity of the observer traveling the geodesics. The travel time along the geodesic $T$ is $T=\frac{b+a\gamma\cot{\frac{m}{2}}}{V+v\gamma\cot{\frac{m}{2}}}$. The second piece connects the two wedges at the points $p_2=(t_{2L}=-T,x_{2L}=-a,y_{2L}=VT)$ and $p_2'=(t_{2L}'=T,x_{2L}'=-a,y_{2L}'=-VT)$. For the two geodesics to form a closed curve, the initial point of one must be the final point of another: $p_1\sim p_2$ and $p_1'\sim p_2'$. In this case the coordinate time $2T$ taken to travel from one wedge to another is balanced exactly by the backward time jump across the wedge between $x_{1L}=a$ and $x_{2L}=-a$. Using (\ref{eq:gott-jump}), we get
\begin{equation}
	\label{eq:gott-closed-cond}
	2T+\Delta t_L=2T+v(x_{2L}-x_{1L})=0,
\end{equation}
or $T=av$, resulting in the relation $V=\frac{b}{av}+\frac{\cot{\frac m2}}{\gamma v}$. The curve is a CTC for $V<1$, which is possible for large enough $a$ whenever $\gamma v>\cot{\frac m2}$. This also sets the lower limit on $a$ for a given boost: $a_{min}=\frac{\gamma b}{\gamma v-\cot{\frac m2}}$ for this choice of geodesics. 

These $x_L=\textrm{const}$ CTCs are not the only ones possible. Cutler \cite{Cutler:1992nn} has determined the (null) boundary of the region containing CTCs, and it turns out that, while there are closed null curves passing closer to the origin than $a_{min}$, no CTCs pass through the origin and all CTCs go counter-clockwise around the origin. 

Moreover, even when CTCs are present, there always exist a neighborhood of each string free of CTCs. As a consequence, ``smoothing out'' the strings in a small enough neighborhood does not affect the CTCs, as noted by Cutler.

\subsection{\label{ssec:gott-unwrap}Unwrapping Gott spacetime}
\begin{figure}
	\centering
	\includegraphics[width=0.9\textwidth]{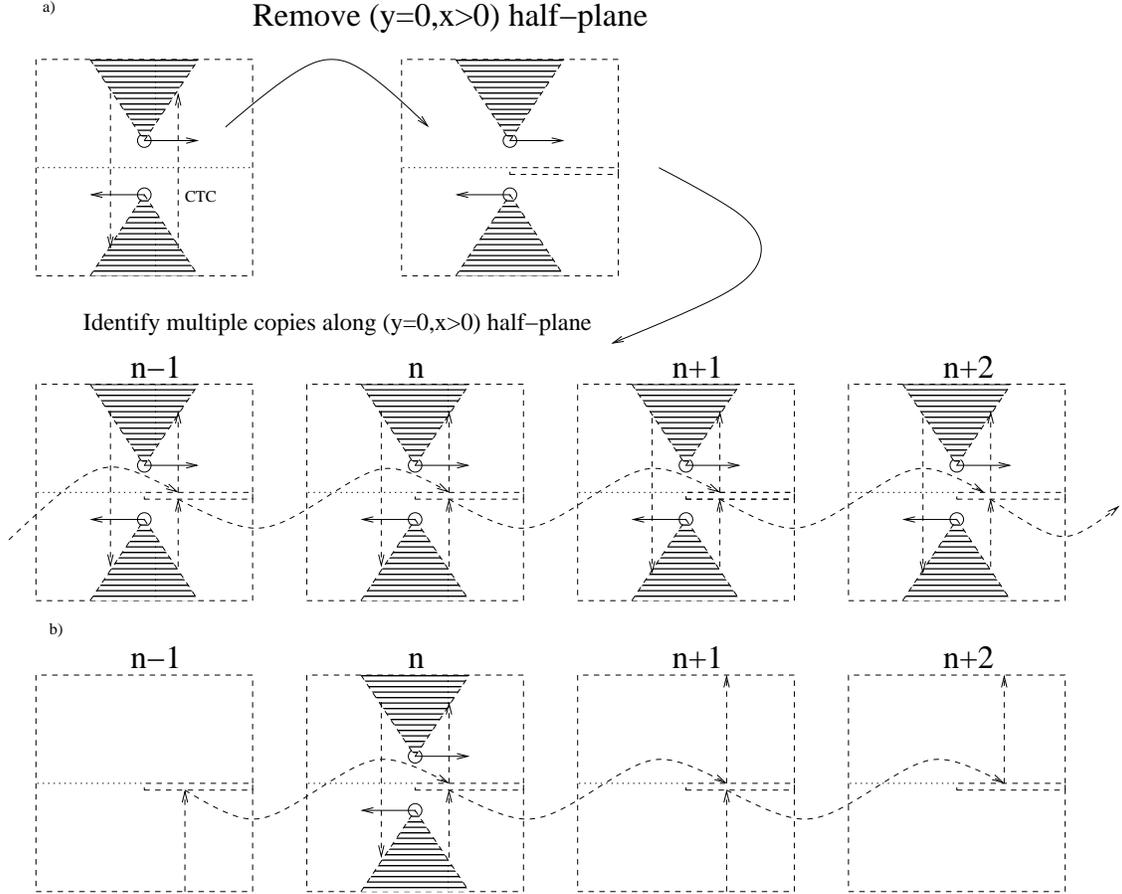}
	\caption{Unwrapping Gott spacetime. a) The $x=y=0$ subspace is removed and the universal cover is constructed by patching multiple Cartesian charts together. The former CTCs (indicated by the vertical arrows) are now open curves passing from chart to chart. The identifications between charts are indicated by the wavy arrows. b) An alternative way to unwrap: a single Gott chart is matched to a collection of Minkowski charts. The former CTCs wrap around just the two strings in the single Gott chart.}
	\label{fig:gott-unwrapped}
\end{figure}

The existence of the minimum value for a CTC's distance from the $x=y=0$ line implies that all CTCs wrap around it in the direction opposite to the relative motion of the strings (they also wrap around both strings at some finite distance from each). We can now follow the same unwrapping procedure as in section \ref{ssec:quasi-regular}: remove the subspace $x=y=0$ and construct the universal covering space of the resulting non-simply connected manifold (we assume that the original Gott spacetime is simply connected, as the strings can be smoothed out). Since in the Gott's choice of identifications neither wedge crosses the $x$-axis, the resulting space can be described using countably many copies of the Cartesian $(t,x,y)$ charts, with the charts $n$ and $n+1$ joined along the non-negative $x$-axis of each one, as shown on Fig. \ref{fig:gott-unwrapped}a. 

This unwrapped Gott spacetime is simply connected, has a quasi-regular singularity at $x=y=0$ and admits no CTCs, no matter how fast the strings are moving relative to each other. It also contains a countable infinity of pairs of boosted strings. The presence of the singularity removes the restriction on the total mass (deficit angle) of all strings, which can now be arbitrarily large, though each string's mass still cannot exceed $\pi$. 

A timelike or null observer starting on the $n$th chart and traveling around both strings counter-clockwise ends up on the $(n+1)$ chart after crossing the positive $x$-axis. This ensures that no such curve is closed, and so no CTCs are present in the unwrapped Gott spacetime. Instead, what used to be CTCs are now open curves winding around two strings per turn. 

It is worth noting that the unwrapped Gott spacetime described above is not the only way to unwrap the CTCs. Since the surface where each two Cartesian charts are joined is locally flat, we do not have to have the two moving strings present on more than one chart. For example, all but one chart can be Minkowski, as shown on Fig. \ref{fig:gott-unwrapped}b. 
 
We have shown that one can indeed change the identifications in the Gott spacetime such that no CTCs are present. The trade-off for the CTCs removal is introduction of a naked quasi-regular singularity. We next turn to another $2+1$ dimensional spacetime with CTCs, the {\G} universe, and demonstrate that a straightforward CTCs unwrapping does not work there.

\section{\label{sec:godel-unwrap}Unwrapping {\G} spacetime}
In this section we construct a spacetime which is locally {\G} at every point, but with a different global structure, such that a given set of CTCs in the original spacetime is not longer closed in the unwrapped spacetime. This is not the only way one can consider when trying to get rid of the {\G} CTCs. One obvious way to remove CTCs from the {\G} space is to restrict the radial coordinate in the chart (\ref{eq:godel-cyl}), thus creating a boundary where the orbits of $\phi$ are still spacelike. An example of this is given in \cite{Brecher:2003rv}, where a preferred holographic screen is constructed at the radial distance $\sinh{r}=1/\sqrt{2}$, where the expansion of the congruence of null geodesics emanating from a point at $r=0$ is zero. Another approach, considered in \cite{Bonnor:1997wz}, is to match the {\G} interior to an exterior spacetime without CTCs. There the metric is explicitly changed locally in the regions where the CTCs used to exist. 

\subsection{\label{ssec:godel-unwrap-def}Unwrapping procedure and circular CTCs}
We now consider a particular example of unwrapping the {\G} space with respect to a given family of CTCs. The metric is given in the $(\tau,r,\phi)$ coordinate chart by the expression (\ref{eq:godel-cyl}). This chart is singular at $r=0$, but this is just a coordinate singularity, as the transformation (\ref{eq:godel-cylind}) shows. Since the spacetime is homogeneous, this chart can be constructed using any point in the spacetime as its origin. The CTCs manifest in this chart are the coordinate curves of $\phi$ when $\sinh(r)>1$. They are not geodesics, but they are isometries of the spacetime, since the metric in this chart does not depend on $\phi$. These CTCs have been extensively studied (see e.g. \cite{Hawking:1973book}). Cooperstock and Tieu in \cite{Cooperstock:2005pu} ``question the continuation of identifying the $\phi$ values of $0$ and $2\pi$ when $\phi$ becomes a timelike coordinate''. Since it is impossible to abruptly change the coordinate identification of $\phi$ only at $\sinh r\ge 1$ without breaking the regularity of the spacetime at $\sinh r=1$, we will instead remove this identification everywhere.

The procedure mirrors the Minkowski case discussed in the section \ref{ssec:quasi-regular}. This is possible because, just like Minkowski spacetime, the {\G} spacetime is everywhere regular, simply connected and is covered by a single global Cartesian chart. We start with the chart (\ref{eq:godel-cartesian}) of the {\G} spacetime $\mc{M}$. In this chart the CTCs that are coordinate curves of $\phi$ cross the positive $x$-axis in the counter-clockwise direction, like the CTCs of the Gott spacetime. We remove the $x=y=0$ subspace (corresponding to a single fiber of the Killing vector $\partial_t$, which also coincides with $\partial_\tau$ at $r=0$) and construct a universal covering space of the resulting non-simply connected manifold $\mc{M}'$. The resulting spacetime can be described by either a single global chart (\ref{eq:godel-cyl}) with $\phi\in\mb{R}$ or by a collection of countably infinitely many Cartesian charts with the charts $n$ and $n+1$ joined along the positive $x$-axis of each one. The description using a single cylindrical chart is possible because the subspace $r=0$, where (\ref{eq:godel-cyl}) is not defined, has been removed, and so this chart is valid everywhere in the unwrapped spacetime.

As expected, the removed subspace makes the unwrapped {\G} space singular, since any curve that passed through the subspace $r=0$ in $\mc{M}$ is incomplete in $\mc{M}'$ and hence in the unwrapped space, as well. Since the b-boundary of a spacetime resulting from unwrapping is singular, one cannot extend the curves through their end points on the b-boundary. 

All circular CTCs cross the positive $x$-axis in $\mc{M}'$ at least once, so their unwrapped-space analogs in the $n$th Cartesian chart end up on the $(n+1)$ chart after crossing the axis and so cannot be closed. This corresponds to the orbits of $\phi$ in the chart (\ref{eq:godel-cyl}) being open in the unwrapped space.

\subsection{\label{ssec:godel-CTCs-shifted}Remaining circular CTCs}
While there are no $r=\textrm{const}>\sinh^{-1}(1)$ CTCs in the unwrapped {\G} spacetime, this is not the only kind of circular CTC present in the {\G} metric. Since the {\G} spacetime is homogeneous, we can construct a cylindrical chart (\ref{eq:godel-cyl}) around any point and obtain CTCs winding around that point. If this new, shifted origin of the cylindrical chart is ``far enough'' from the old, unshifted one, then the CTCs around it will lie wholly on a single sheet of the space unwrapped around the unshifted origin, and so will remain CTCs even in the unwrapped space.

To demonstrate this, it is convenient to use the quotient space $\tilde{\mc{M}}$ of the $(2+1)$-dimensional {\G} space $\mc{M}$ with the fiber defined by the orbits of the timelike Killing vector $y$: $y:\mc{M}\to\tilde{\mc{M}}$. The metric $h_{ab}$ of the quotient space is calculated as $h_{ab}=g_{ab}-\frac{y_a y_b}{y^c y_c}$ (see e.g. \cite{Geroch:1970nt}). Since $y^a$ is timelike everywhere, with a non-vanishing norm, $h_{ab}$ is nowhere singular, the reduced space is Riemannian, and its line element in the original Cartesian coordinates is just a flat two-dimensional space
\begin{equation}
	\label{eq:godel-reduction-y}
	ds^2=dx^2+dt^2.
\end{equation}

An arbitrary circular CTC in the full space, parameterized by $(0\le\phi\le 2\pi)$, is completely defined by its center $(t_0,x_0,y_0)$ and radius $R$. It can be written using the equivalent of (\ref{eq:godel-cylind}) as
\begin{equation}
\begin{split}
	\label{eq:godel-cylind-shift}
	e^{x-x_0}&=\cosh2R+\cos\phi \sinh 2R \\
	(y-y_0)e^{x}&=\sqrt{2}\sin\phi \sinh 2R \\
	\tan\frac 12 \left(\phi+\frac{t-t_0}{\sqrt{2}}\right)&=e^{-2R}\tan\frac 12 \phi.
\end{split}
\end{equation}

An image of this CTC in the flat quotient space $\tilde{\mc{M}}$ is obtained by omitting the coordinate $y$ from (\ref{eq:godel-cylind-shift}):
\begin{equation}
\begin{split}
	\label{eq:godel-cylind-flat}
	e^{x-x_0}&=\cosh2R+\cos\phi \sinh 2R \\
	\tan\frac 12 \left(\phi+\frac{t-t_0}{\sqrt{2}}\right)&=e^{-2R}\tan\frac \phi 2,
\end{split}
\end{equation}
which can be rewritten in an explicit form as
\begin{equation}
\begin{split}
	\label{eq:godel-cylind-flat-explicit}
	x&=x_0+\ln{\cosh2R+\cos\phi \sinh 2R} \\
	t&=t_0+2\sqrt{2}\tan^{-1}\frac{(e^{-2R}-1)\tan\frac \phi 2}{1+e^{-2R}\tan^2\frac \phi 2}.
\end{split}
\end{equation}
All CTCs with the same values of $x_0$ and $t_0$, but with a different $y_0$ are mapped into the same closed curve. The image of the b-boundary of the unwrapped space is the point $x=t=0$ of the quotient space. 

\begin{figure}
	\centering
	\psfrag{x}{\small{$\frac{x}{2R}$}}
	\psfrag{t}{\small{$\frac{t}{2\sqrt{2}\tan^{-1}\sinh R}$}}
	\psfrag{R=Rmin}{\small{$R=R_{min}$}}
	\psfrag{R=5}{\small{R=5}}
	\psfrag{R=20}{\small{R=20}}
	\includegraphics[width=1.0\textwidth]{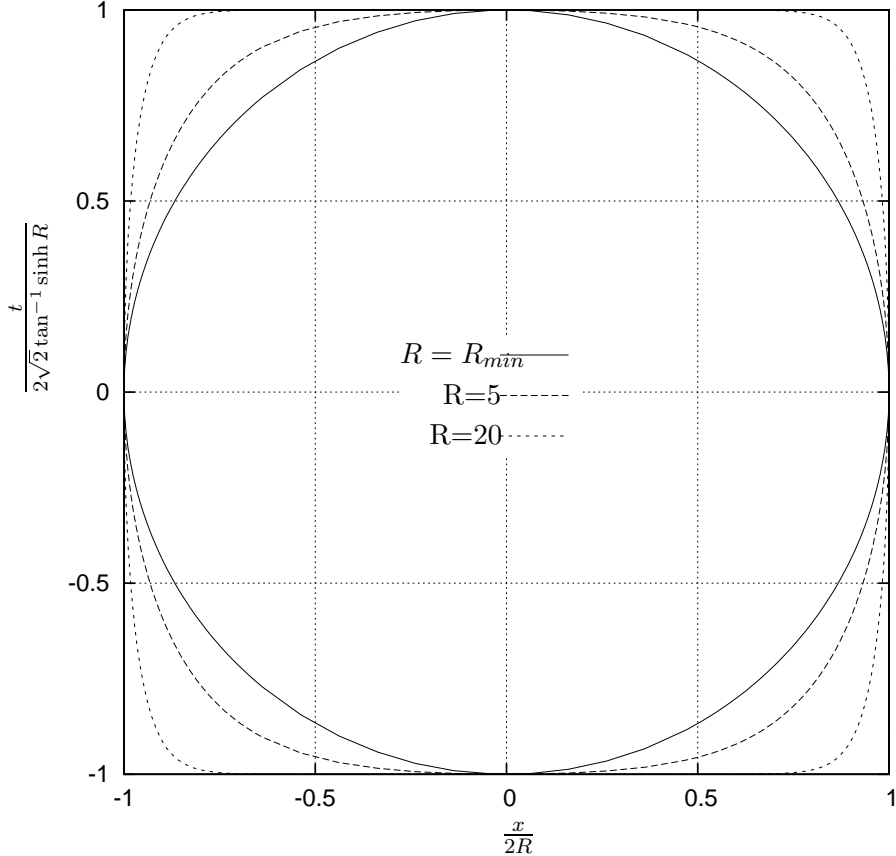}
	\caption{Normalized concentric CTCs of radius $R$ in the $(x/2R,t/(2\sqrt{2}\tan^{-1}\sinh R))$ coordinates. The CTCs lie inside a square with the side equal to two, but outside of a circle inscribed into it.}
	\label{fig:ctcs-sector-inscribed}
\end{figure}

We note the following properties of a curve described by (\ref{eq:godel-cylind-flat-explicit}): 
\begin{enumerate}
\item \label{godel-inscribed-1}It lies wholly inside a rectangle centered at $(x_0,t_0)$ and with sides $2R$ and $2\tan^{-1}\sinh{R}$, and 
\item \label{godel-inscribed-2}It lies wholly outside of an ellipse inscribed into this rectangle, as described by the equation (\ref{eq:godel-inscribed-ctc}).
\end{enumerate}
\begin{equation}
	\label{eq:godel-inscribed-ctc}
	\left(\frac{x-x_0}{2R}\right)^2+\left(\frac{t-t_0}{2\sqrt{2}\tan^{-1}\sinh R}\right)^2=1.
\end{equation}
The property \ref{godel-inscribed-1} follows from the ranges of $x$ and $t$ in (\ref{eq:godel-cylind-flat-explicit}) for a given $R$, while the property \ref{godel-inscribed-2} can be shown by substituting (\ref{eq:godel-cylind-flat-explicit}) into the left-hand side of (\ref{eq:godel-inscribed-ctc}), finding the four minima of the resulting function of $\phi$ and showing that they give exactly the equality (\ref{eq:godel-inscribed-ctc}). At small $R$ the curve (\ref{eq:godel-cylind-flat-explicit}) tends closer to (\ref{eq:godel-inscribed-ctc}), while at large $R$ it asymptotically approaches the rectangle, as shown on Fig.\ref{fig:ctcs-sector-inscribed}.

We can now show that there exist CTCs that are not unwrapped by the singular b-boundary resulting from unwrapping around the origin in the chart (\ref{eq:godel-cyl}). Since the image of any CTC winding around the b-boundary has to wrap around the image of the b-boundary in the reduced space, constructing the image of a CTC that does not wrap the image of the b-boundary is enough to show the existence of CTCs that are not unwrapped. Since the image of any CTC of radius $R$ with, say, $|x_0|>2R$ or $|t_0|>2\sqrt{2}\tan^{-1}\sinh R$ does not wrap around $x=t=0$, all such curves remain closed after unwrapping.

Thus the naive coordinate identification change of \cite{Cooperstock:2005pu} fails to unwrap at least some of the circular CTCs.

\section{\label{sec:godel-unwrap-multi}Multiple unwrapping of the {\G} space}
By unwrapping the {\G} spacetime we have introduced a quasi-regular singularity into it, yet we did not accomplish the goal of removing all CTCs from it. If one remains intent on also removing circular CTC, one may consider the unwrapping in several charts at once instead of just one. 

\subsection{\label{ssec:double-unwrapping}Double Unwrapping}
We first consider what happens when we unwrap the CTCs in just two charts. We show that even in this simple case the unwrapped space contains a countably infinite number of quasi-regular singularities.
 
This``double unwrapping'' procedure would look as follows. We consider two different families of circular CTCs, one centered at $x=x_1,y=y_1$ and $x=x_2,t=t_2$ respectively. The subspaces $(x=x_1,t=t_1)$ and $(x=x_2,t=t_2)$ are the fixed points of the corresponding $U(1)$ isometries represented by the CTCs. As before, we first remove these fixed points from the spacetime $\mc{M}$, only this time we have to remove both sets, resulting in the singular space $\mc{M}^{\prime\prime}$. Next we construct the universal covering space $\bar{\mc{M}}$ of $\mc{M}^{\prime\prime}$ and lift the metric tensor to $\bar{\mc{M}}$. Since the original $(2+1)$-{\G} manifold is $\mb{R}^3$, we do not need to worry about accidentally removing the topological features unrelated to unwrapping. In this case $\mc{M}^{\prime\prime}$ is homotopy-equivalent to the wedge sum of two circles $\mb{S}^1\vee\mb{S}^1$, known as the ``figure 8'' (see e.g. \cite{Hatcher:2001book}, chapter 1). This can be shown by explicitly constructing a deformation retraction, an operation that preserves the fundamental group of a manifold. To do that, we first note that, since the orbits of $y$ are open lines, they can be retracted into points by the continuous map $f_s:(t,x,y)\to(t,x,(1-s)y)$. $f_0$ is the identity map, and $f_1$ maps $\mc{M}^{\prime\prime}$ into a two-dimensional plane with two points ($(x_1,t_1)$ and $(x_2,t_2)$) removed. Following \cite{Hatcher:2001book}, we next retract the plane first onto two circles (one around each removed point) connected by a line segment, then contracting the connecting segment into a point. The resulting space is the``figure 8''. As a result of the retraction, the singularities now "fill the inside of the circles". The fundamental group of $\mc{M}^{\prime\prime}$ is the fundamental group of the``figure 8", which is just the free product of two copies of $\mb{Z}$, $\pi_1=\mb{Z}*\mb{Z}$. Each element of the group corresponds to winding around one of the two singularities in $\mc{M}^{\prime\prime}$.

\begin{figure}
	\centering
	\includegraphics[width=1.0\textwidth]{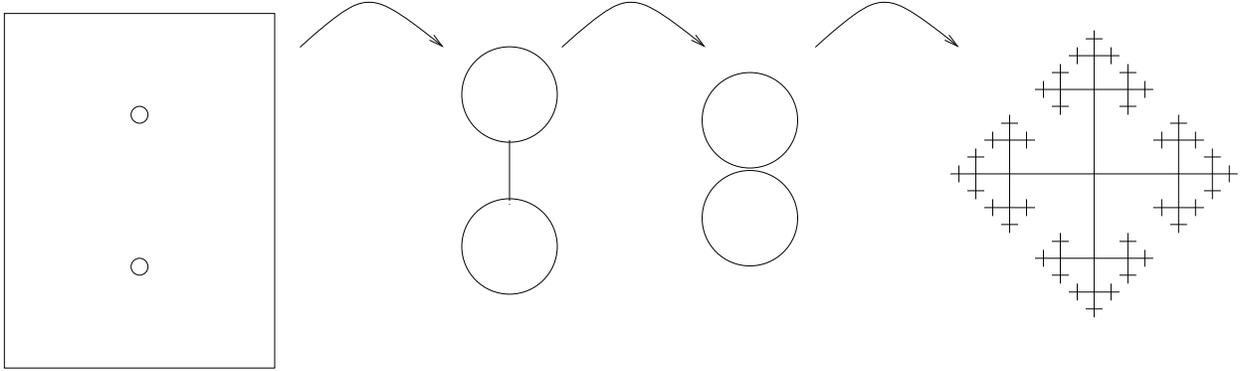}
	\caption{Constructing the homotopy equivalence of the twice-unwrapped {\G} space. We start with the $(t,x,y=0)$ subspace with two points removed (left), then retract the space onto first two circles connected by a line segment, then the ``figure 8'', and finally construct the universal cover of the ``figure 8'' (only the first four levels of nodes are shown). Each edge corresponds to a circle wrapping around one of the two singularities in the original space, so the unwrapped space contains infinitely many singularities.} 
	\label{fig:fig8-cover}
\end{figure}

The universal cover of the``figure 8'' is well known, it is a tree with countably infinitely many edges and each node connecting four edges (see e.g. \cite{Hatcher:2001book} for construction). The process of constructing the universal cover of the twice punctured plane is shown schematically on Fig. \ref{fig:fig8-cover}. 
Each edge of this graph corresponds to a CTC winding around one of the removed subspaces of fixed points. In the unwrapped space this CTC becomes open and corresponds to a given path along the graph. Traversing one edge corresponds to ``going around'' one of the singularities, so there is a one-to-one correspondence between the singularities in the twice-unwrapped {\G} and the edges in the graph.

We can now conclude that unwrapping around two axes at once in a simply connected spacetime results in a spacetime with a countable infinity of quasi-regular singularities of the type discussed in section \ref{ssec:quasi-regular}.
 
\subsection{\label{ssec:multiple-unwrapping}Multiple Unwrapping}
As discussed in section \ref{ssec:godel-CTCs-shifted}, there are infinitely many families of concentric CTCs parameterized by $(t_0,x_0,y_0)$ for which at least some CTCs persist after unwrapping around $(x=0,t=0)$. According to the property \ref{godel-inscribed-2}, each CTC lies outside an ellipse (\ref{eq:godel-inscribed-ctc}). All such ellipses are larger than the image of a closed null curve corresponding to $\sinh{R}=1$. If we modify the spacetime in a way that transforms any such ellipse into an open curve, the resulting spacetime will not have any circular CTCs. This can be accomplished by first tessellating the quotient space $(x,t)$ with a dense enough triangular lattice, such that there is a vertex inside any ellipse with $\sinh{R}=1$, then lifting it into the full 3-dimensional {\G} spacetime using $y$ as the fiber and finally by going to the universal covering space, in a procedure analogous to the one described in section \ref{ssec:double-unwrapping}. 

\begin{figure}
	\psfrag{x}{\small{$\frac{x}{2R}$}}
	\psfrag{t}{\small{$\frac{t}{2\sqrt{2}\tan^{-1}\sinh R}$}}
	\centering
		\includegraphics[width=0.5\textwidth]{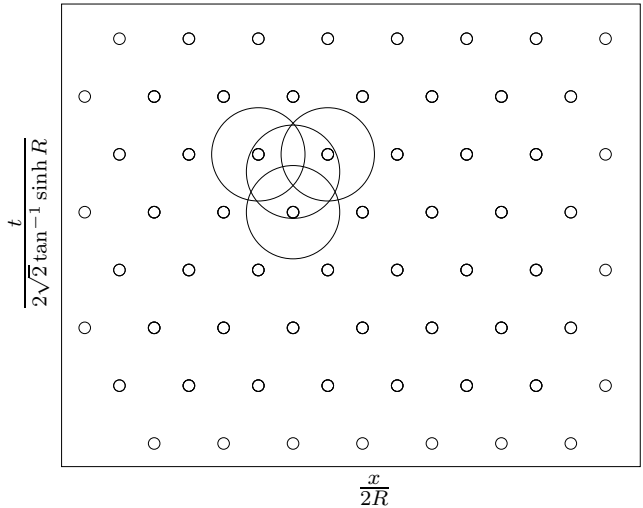}
	\caption{A tessellation of the two-dimensional quotient space of the {\G} spacetime. Each vertex corresponds to a timelike line in the full spacetime and each circle is an image of a closed null curve $R=R_{min}$. The tessellation is dense enough to make any circular CTCs wrap around at least one such line, as shown. Once the lines are removed and the resulting non-simply connected spacetime is lifted into the full spacetime and then into its universal cover, no circular CTCs are present in this ``multiply unwrapped'' spacetime.} 
	\label{fig:godel-tessel}
\end{figure}

\begin{figure}
	\centering
		\includegraphics[width=0.5\textwidth]{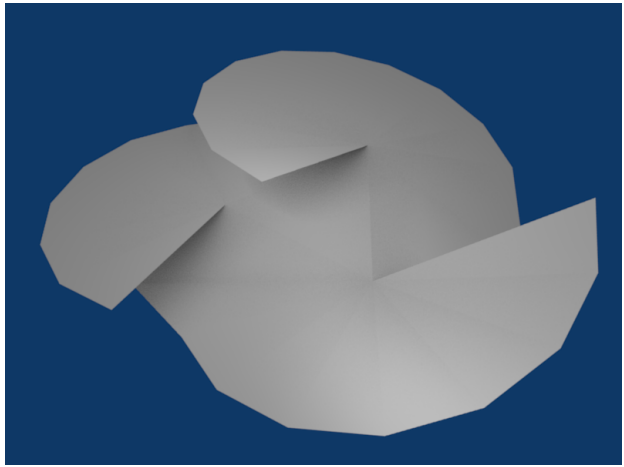}
	\caption{A visualization of the two-dimensional quotient space of the {\G} spacetime unwrapped at three points at once. Only a small patch of this space is shown, as unwrapping around two or more points results in a countable infinity of singular b-boundaries.} 
	\label{fig:triple_helix}
\end{figure}

The tessellation is shown schematically on Fig. \ref{fig:godel-tessel}.
The fundamental group of the tessellated space is the free group on $\mb{Z}$ generators, one for each removed point. A small patch of the unwrapped quotient space is illustrated on the Fig. \ref{fig:triple_helix}. Each helix corresponds to a family of unwrapped concentric circular CTCs. 
The price to pay for removing all circular CTCs is the introduction of a naked singular b-boundary consisting of a countable infinity of disjoint pieces.

\subsection{\label{ssec:godel-sector}Sector-like CTCs in the {\G} space}
One can ask whether any other types of CTCs are present in the multiply unwrapped {\G} spacetime. For example, is it possible to weave one's way in between the vertices and come back to the starting point along a CTC? This seems unlikely and we conjecture that no such CTCs exist. To support this conjecture we describe a different kind of CTCs, we call sector-like CTCs, and show that they, too, are transformed into open curves by the multiple unwrapping procedure of the section \ref{ssec:multiple-unwrapping}.

The idea of constructing a CTC surviving the tessellation of the section \ref{ssec:multiple-unwrapping} is to exploit the property of the {\G} spacetime where an arc with a larger radius but a smaller angular distance can get us just as far back in time in the chart (\ref{eq:godel-cyl}). The hope is then to go far along a timelike curve in a radial direction, then along an arc, then back to the starting point, thus covering a sector instead of a full circle in this chart. If the resulting sector is thin enough, then we can try to fit it in the tessellated space in such a way that no vertices are inside the sector. 

To check if this can be done, we calculate the angular and linear distance along the arc required to overcome the time lost traveling forward and back along the two radial directions. Since a closed null sector would be ``thinner'' than the corresponding timelike sector, we analyze the null sector first.

The (negative) time change in the coordinate $\tau$ along a null arc of radius $R$ and angle $\Delta\phi$ in the chart (\ref{eq:godel-cyl}) can be calculated as 
\begin{equation}
	\label{eq:godel-arc-tau-jump}
	\Delta\tau=(\sinh^2R(\sqrt{2}-\coth{R}))\Delta\phi,
\end{equation}
and it has to compensate for the positive time change of $2R$ along the two radii of the sector, resulting in the total angular change of
\begin{equation}
	\label{eq:godel-arc-ang-jump}
	\Delta\phi=\frac{2R}{\sinh^2R(\sqrt{2}-\coth{R})}.
\end{equation}

\begin{figure}
	\centering
		\includegraphics[width=1.0\textwidth]{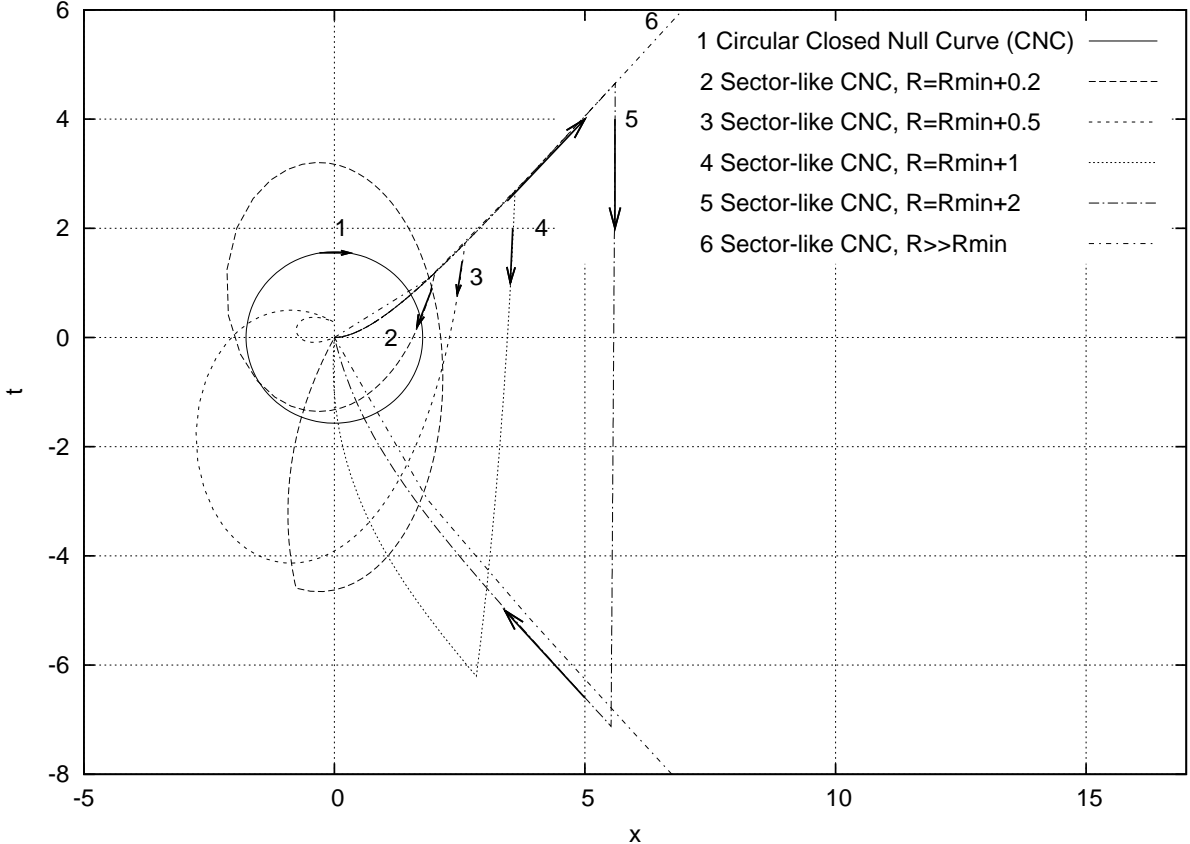}
	\caption{Images of sector-like closed null curves in the $(x,t)$ coordinates for a range of values $R$. Arrows indicate the future null directions. For large $R$ the curve tends to a triangle with a fixed minimum angle. The angle increases as $R$ goes down, and for $R$ close to the minimum value of $\sinh{R_{min}}=1$ the null curve has to wrap around the origin several times to compensate for the time lost along the radial paths.} 
	\label{fig:ctcs-sector}
\end{figure}

To see if the sector is thin enough to fit between the vertices of the lattice for large enough $R$, we project it into the flat quotient space (\ref{eq:godel-reduction-y}). The three legs of the path written in the $(\tau,r,\phi)$ coordinates and parameterized by $\lambda$ are
\begin{equation}
\begin{split}
	\label{eq:godel-sector-null-polar}
	&\left(\lambda,\lambda,0\right), 0\le\lambda\le R,\\
	&\left(2R-\lambda,R,\frac{\lambda-R}{\sinh^2R(\sqrt{2}-\coth{R})}\right), R\le\lambda\le 3R,\\
	&\left(\lambda-4R,4R-\lambda,\frac{2R}{\sinh^2R(\sqrt{2}-\coth{R})}\right), 3R\le\lambda\le 4R.
\end{split}
\end{equation}
The same curve in the $(x,t)$ chart of (\ref{eq:godel-reduction-y}) can be described using the explicit coordinate transformation (\ref{eq:godel-cylind-flat-explicit}).
The resulting curves are plotted for several values of $R$ on Fig.\ref{fig:ctcs-sector}.
For large $R$ the three turning points of the path in the $(x,t)$ chart asymptotically approach $(0,0), (2R,2\sqrt{2}R), (2R,-2\sqrt{2}R)$. Thus, no matter how large $R$ is, the tessellation dense enough to unwrap circular CTCs also unwraps the sector-like closed null (and therefore timelike) curves. In this sense, the original CTCs described by {\G} appear to have the smallest ``footprint'' in the flat quotient space.

Whether or not there are other CTCs that persist in the multiply unwrapped spacetime, it is quite clear that removing CTCs from the {\G} spacetime solely by changing the coordinate identification results in a rather contrived space with a countable infinity of naked quasi-regular singularities.

\section{\label{sec:conclusion}Conclusion}
We have defined and investigated "unwrapping" CTCs in two $(2+1)$-dimensional toy models, the {\G} universe and the Gott spacetime, as a concrete implementation of the claim by Cooperstock and Tieu \cite{Cooperstock:2005pu} that the periodic identification of a timelike coordinate is "purely artificial". The procedure requires removing a timelike line from the spacetime and constructing a universal cover of the resulting non-simply connected spacetime. We have demonstrated that such an unwrapping creates a naked quasi-regular singularity, corresponding to the removed timelike line in the original space. The same argument can be extended to other $(2+1)$-dimensional spacetimes where the CTCs follow a periodically identified coordinate, as long as the coordinate curves are contractible to a fixed point.

While the unwrapped Gott spacetime is devoid of CTCs, the unwrapped {\G} spacetime still contains them. We have defined a "multiple unwrapping" of the {\G} spacetime in order to remove the remaining circular CTCs. As a result, this multiply unwrapped spacetime contains a countably infinite number of singularities. We conjecture that this multiple unwrapping removes all other CTCs as well, and support it by giving an explicit example of a sector-like CTC, which is also removed by the multiple unwrapping. 

Our investigation into the ways of removing the CTCs by means of changing coordinate identifications suggests that CTCs appearing in the solutions of the Einstein equation are not an mathematical artifact of arbitrary coordinate identifications, but rather an unavoidable, if undesirable, consequence of General Relativity, provided only that we do not introduce naked quasi-regular singularities.

\begin{acknowledgments}
The author would like to thank Kristin Schleich, Don Witt and Kory Stevens for helpful discussions, suggestions and observations. This work was sponsored in part by a UBC UGF award.
\end{acknowledgments}

\end{document}